# Multiscale Mechanical Response of 3D-Printed Diamondiynes: From Movable Interlocked Lattices to Architected Metamaterials

Anitesh Kumar Singh[1†], Rodrigo A. F. Alves[2†], Tapas Pal[3], Sarmistha Bora[4], Hugo X. Rodrigues[2], Emanuel J. A. dos Santos[2], Camila de L. Ribeiro[5], Alysson M. A. Silva[5], Luiz A. Ribeiro Júnior[2], Douglas S. Galvão[6*], Chandra Sekhar Tiwary[1*]

[1] Department of Metallurgical and Materials Engineering, Indian Institute of Technology Kharagpur, WB, India-721302

[2] Computational Materials Laboratory, LCCMat, Institute of Physics, University of Brasília, 70910-900 Brasília, Federal District, Brazil.

[3] School of Nano Science and Technology, Indian Institute of Technology Kharagpur, WB, India-721302

[4] Department of Chemistry, Cotton University, Assam, India-781001

[5] University of Brasília, College of Technology, Department of Mechanical Engineering, 70910-900, Brasília, Federal District, Brazil.

[6] Department of Applied Physics and Center for Computational Engineering and Sciences, State University of Campinas, Campinas, 13083-859, SP, Brazil

* Correspondence: galvao@ifi.unicamp.br, chandra.tiwary@metal.iitkgp.ac.in

**Abstract:** Diamondynes are a recently synthesized three-dimensional carbon allotrope, with interlocked and movable sublattices that introduce deformation modes not present in standard architected materials. Here, we report the first multiscale mechanical assessment of Diamondiyne-derived architectures by combining quasi-static compression of 3D-printed specimens with reactive molecular dynamics simulations of the corresponding atomic-scale models. We generate four geometries (3F, 2F-SY, 4F, and 2F-USY). All structures resulted in lower density in the range of 0.20-0.38 g·cm$^{-3}$. Experiments indicate that the symmetric two-sublattice structure (2F-SY) delivers the best performance, reaching a specific yield strength of 5.91 MPa·g$^{-1}$·cm³ and a specific energy absorption of 279 J·g$^{-1}$, whereas 2F-USY architecture yielded the lowest values, with 0.77 MPa·g$^{-1}$·cm³ and 16 J·g$^{-1}$. The 4F geometry provided a specific energy absorption of 254 J·g$^{-1}$. The structures deformed through geometric collapse and strut buckling, which was due to diagonal shear in 2F-USY and progressive compaction in 2F-SY and 3F. Molecular dynamics simulations also confirmed these experimental trends and revealed strong directional anisotropy due to the arrangement of interlocked sublattices, with a stiffness of 24.1 GPa along the z-direction in the case of 4F architecture. Overall, Diamondiyne-derived architectures display geometry-dominated mechanical behavior and serve as a promising platform for lightweight, energy-absorbing metamaterials.

# 1. Introduction

Three-dimensional carbon allotropes [1] have recently emerged as an attractive platform for research on structural and functional materials due to their broad tunability [2], intrinsic porosity [3], and exceptional stiffness-to-weight ratios [4] [5]. Architected carbon networks, such as schwarzites [6], gyroids [7], pentadiamonds [8] [9], and nanotube-based lattices [10] have demonstrated that the mechanical response in open frameworks is primarily governed by topology rather than by chemical composition. Importantly, this particular trend enables geometry-driven control of stiffness, anisotropy, structural failure modes, and energy absorption [11]. This paradigm has motivated the design of novel metamaterials [12] in which structural performance can be engineered from atomic-scale motifs and reliably reproduced through additive manufacturing, revealing a remarkable level of scale-independent mechanical behavior. Among the newest additions to the family of three-dimensional carbon networks is Diamondiyne [13] [14], a recently synthesized allotrope composed of covalently bonded –C≡C–C≡C– linkers arranged in a porous lattice. High-resolution iDPC-STEM imaging has revealed that Diamondiyne exhibits asymmetric two-fold interpenetration between independent carbon frameworks, an architectural motif not previously observed in known carbon phases [13] [14]. Subsequent theoretical work demonstrated that these interpenetrated frameworks are not covalently connected and may undergo relative motion within the crystal, giving rise to a class of movable interlocked lattices [13]. This implies the existence of internal degrees of freedom that have no precedent in crystalline carbon and raises the question of how such structural mobility influences mechanical performance across different scales. Despite this unique topology, the mechanical behavior of Diamondiynes and their related interlocked architectures remains unexplored, both experimentally and through multiscale simulations. Previous studies on architected materials derived from schwarzites [6], tubulanes [15], pentadiamonds [8], and carbon nanotube networks [10] have shown a strong correspondence between atomistic and macroscale deformation. It is worth noting that none of these systems possesses movable, interpenetrating sublattices. Therefore, the influence of sublattice mobility, symmetry, and multiplicity on mechanical response is unknown and constitutes a significant gap in the current literature.

Printing these structures needs extra care because the struts are delicate and the geometry is complex. Additive manufacturing, which works through layer-by-layer deposition [16] and can deposit complex structures [17] with lower scrap [18], is one of the most effective techniques for creating these complex diamondyne architectures. Fused deposition modeling (FDM) is a well-established technique for printing polymers [19]. Singh et al. [6] [20] printed Schwarzites- and Schwarzynes-inspired Polylactic Acid (PLA) and Thermoplastic Polyurethane (TPU) based structures using FDM, and analyzed how to reduce stress concentration and enhance structural performance in another study by Ambekar et al. [21], PLA-based schwarzite structures were fabricated using FDM to analyze the impact behavior of the printed structure. Previous studies confirmed the FDM technique as one of the most used processes to print complex polymer structures [22] [23].

Herein, we carried out the first multiscale mechanical investigation of Diamondiyne-derived architectures, combining reactive molecular dynamics simulations with quasi-static compression tests on upscaled structures fabricated by FDM. Four architectures (2F-SY, 2F-USY, 3F, and 4F [14]) were generated by geometrically replicating unit-cell atomistic models while preserving the relative arrangement of their interlocked sublattices. This approach enables a direct evaluation of how topological features unique to Diamondiynes, including sublattice symmetry and the number of interpenetrated frameworks, determine their nanoscale and macroscale mechanical behavior.

## 2. Methodology

### 2.1. Experimental details

As mentioned above, Diamondiyne-derived architectures were fabricated by geometrically replicating the unit-cell atomistic models of the 2F-SY, 2F-USY, 3F, and 4F frameworks. A standard slicing workflow was used to prepare the macro-scale geometrical models for additive manufacturing, which were exported in STL format. Fused deposition modeling (FDM) [19], a material-extrusion additive manufacturing (3D-printing) technique, was used to fabricate the Diamondiyne structures. This technique works by melting the thermoplastic (PLA in the present study) filament and sequentially depositing layers according to the toolpath specified in an STL file. A schematic of the FDM process used in the present study is presented in **Fig. 1**.

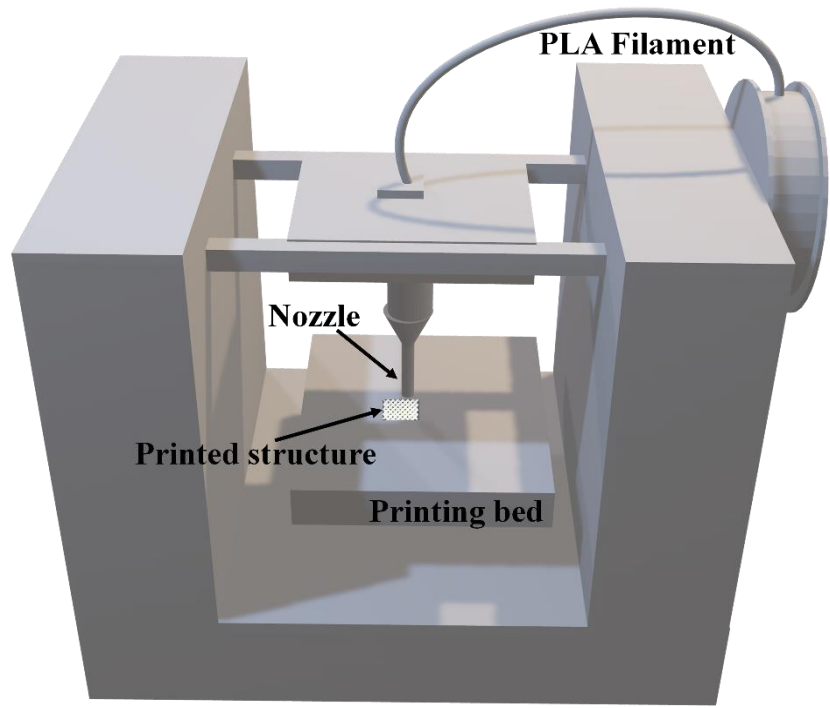

**Fig. 1**. Schematics of the FDM with a single nozzle PLA.

Four Diamondiyne-based architectures (3F, 2F-SY, 4F, and 2F-USY) with three different nominal sizes (15 mm, 20 mm, and 25 mm), and with three replicates of each structure were printed, which resulted in cubic or near-cubic 3D structures, as shown in **Fig. 2**. A nozzle temperature of 200 °C and a printing speed of 35 mm·s$^{-1}$ was programmed to ensure smooth flow and defect-free deposition of melted PLA on a 60 °C preheated bed. The deposition was 100% infill with a layer height of 0.2 mm [6].

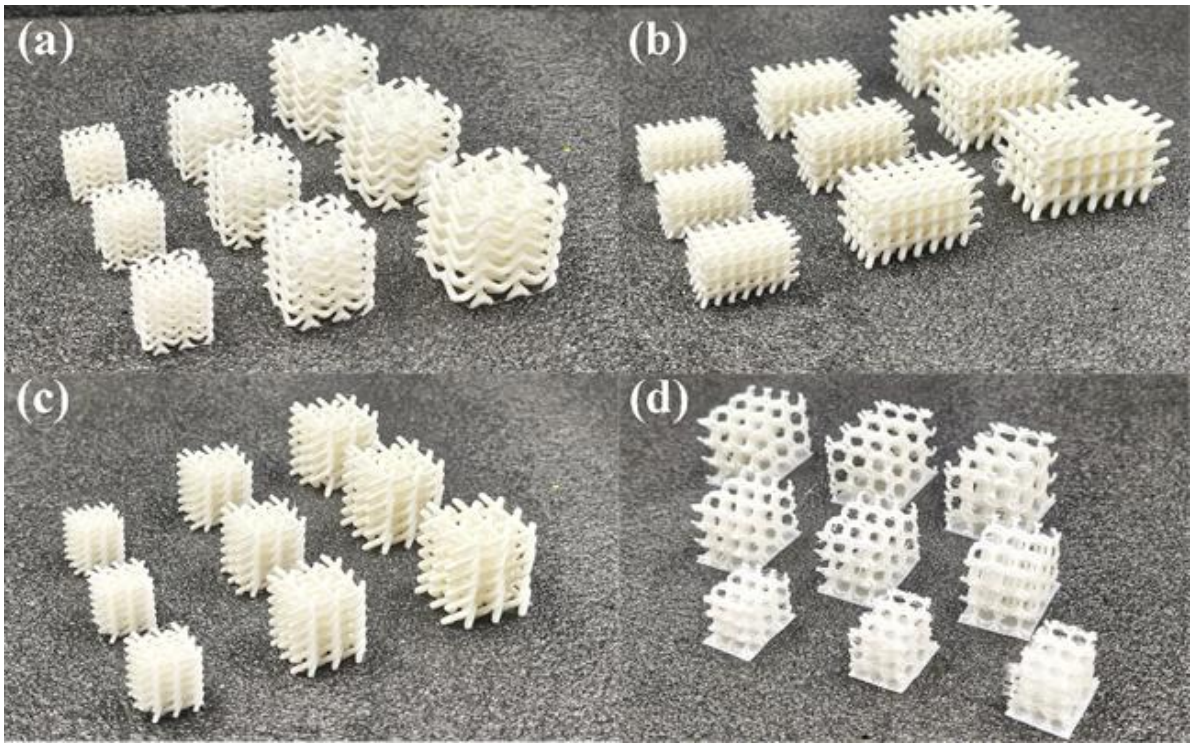

**Fig. 2.** 3D-printed Diamondiyne-derived architectures used in this study: (a) 3F-DCS, (b) 2F-SY-DCS, (c) 4F-DCS, (d) 2F-USY-DCS, each printed at three different nominal sizes (15, 20, and 25 mm).

The 3D-printed structures, including dimensions and sample notations, are summarized in **Table 1**. Three sizes, each with three replicates per structure type, were 3D-printed using FDM to evaluate geometric scaling effects under uniaxial compression. The room-temperature compression test was performed on each structure using an Instron machine with a maximum load capacity of 5 kN. The compression was done at a constant compression rate of 0.5 mm·s$^{-1}$. Visual monitoring during the compression test was conducted using an optical camera. The compressive stress was evaluated by dividing the applied force by the initial cross-sectional area, while strain was obtained from the actuator displacement normalized by the specimen height. The density of the structure was calculated by dividing its mass by its volume.

By summing the stress response up to the densification point, we determined the yield strength, resilience, and energy absorption from the stress-strain curves. To ensure the study's reliability, the results were adjusted to their own units by dividing by the density of the structure to which they were related. Additionally, a post-mortem examination of broken specimens was conducted to determine how they bent, buckled, and structurally failed in ways unique to each architecture.

**Table 1.** Dimensions and sample notation for all Diamondiyne-derived architectures printed in this study.

| Structure Type | Dimension (mm) | Sample Notation |
|---|---|---|
| 3F-DCS | 15×15×16.08 | S1-15 |
| | 20×20×21.44 | S1-20 |
| | 25×25×26.8 | S1-25 |
| 2F sym-DCS | 15×15×25.2 | S2-15 |
| | 20×20×33.5 | S2-20 |
| | 25×25×42 | S2-25 |
| 4F | 15×15×15 | S3-15 |
| | 20×20×20 | S3-20 |
| | 25×25×25 | S3-25 |
| 2F unsym-1-DCS | 15×15×20 | S4-15 |
| | 20×20×25 | S4-20 |
| | 25×25×33.3 | S4-25 |

**2.2. Computational methods**

We used the LAMMPS package [22] to perform fully atomistic reactive molecular dynamics simulations to investigate how Diamondiyne-derived architectures respond to nanoscale stress. The previously optimized Diamondiyne structures were used to make the atomistic models of the 2F-SY, 2F-USY, 3F, and 4F frameworks [14] (see **Fig. 3**), preserving the relative arrangement, symmetry, and interpenetration of their sublattices. All systems were modeled as fully periodic simulation cells to avoid surface effects and to ensure homogeneous stress distribution during deformation.

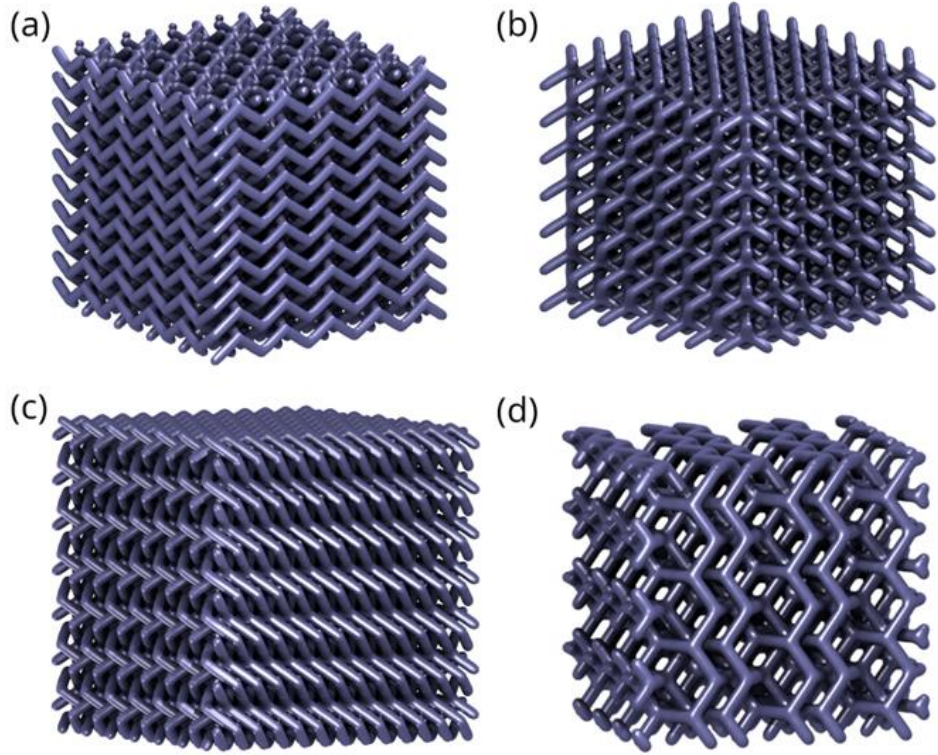

**Fig. 3.** Atomistic models of the four simulated structures: (a) 3F – S1, (b) 2F-SY – S2, (c) 4F – S3, and (d) 2F-USY – S4. These configurations were used as the initial geometries for the mechanical simulations under uniaxial compression.

Interatomic interactions were described using the ReaxFF reactive force field [24] [25], which enables bond formation and breaking throughout the simulation. A time step of 0.25 fs was used to accurately describe the dynamics of highly strained sp–sp²–sp³ hybridized carbon networks. Before being loaded, each structure was equilibrated for 200 ps in the isothermal–isobaric ensemble (NPT) at 300 K and 1 atm using Nosé–Hoover thermostats and barostats[26] [20] with damping constants of 100 fs and 1000 fs, respectively. This process ensured that the temperature remained stable, the stress was relieved, and any residual forces were eliminated. Uniaxial compression simulations were then performed in the canonical ensemble (NVT) at a constant engineering strain rate of $10-5$ fs⁻¹ along the loading direction, with the lateral dimensions allowed to relax. Three independent directions (x, y, and z) were evaluated to capture anisotropy arising from sublattice arrangement. Engineering stress was computed from the virial stress tensor normalized by the instantaneous cross-sectional area, and engineering strain was calculated based on the deformation of the simulation box.

To enable direct comparison with experiments, the stress–strain curves were used to extract elastic modulus, yield point, and energy absorption up to the onset of structural densification. Visualization of deformation pathways and local stress concentrations was carried out using the Visual Molecular Dynamics (VMD) software [27], which facilitated correlation between atomistic mechanisms and macroscale behavior. All simulations were run for at least 500 ps or until structural collapse.

# 3. Results and discussion

## 3.1. Experimental stress-strain response

We begin by presenting the results of the compressive stress–strain response of the four Diamondiyne-derived architectures for the three printed sizes (see **Fig. 3**). All structures exhibit the characteristic nonlinear behavior associated with bending-dominated architected lattices, featuring an initial elastic regime, followed by progressive collapse of the porous domains and a final densification stage. Despite these standard features, each architecture displays a distinct mechanical signature governed by the number, arrangement, and symmetry of its interlocked sublattices.

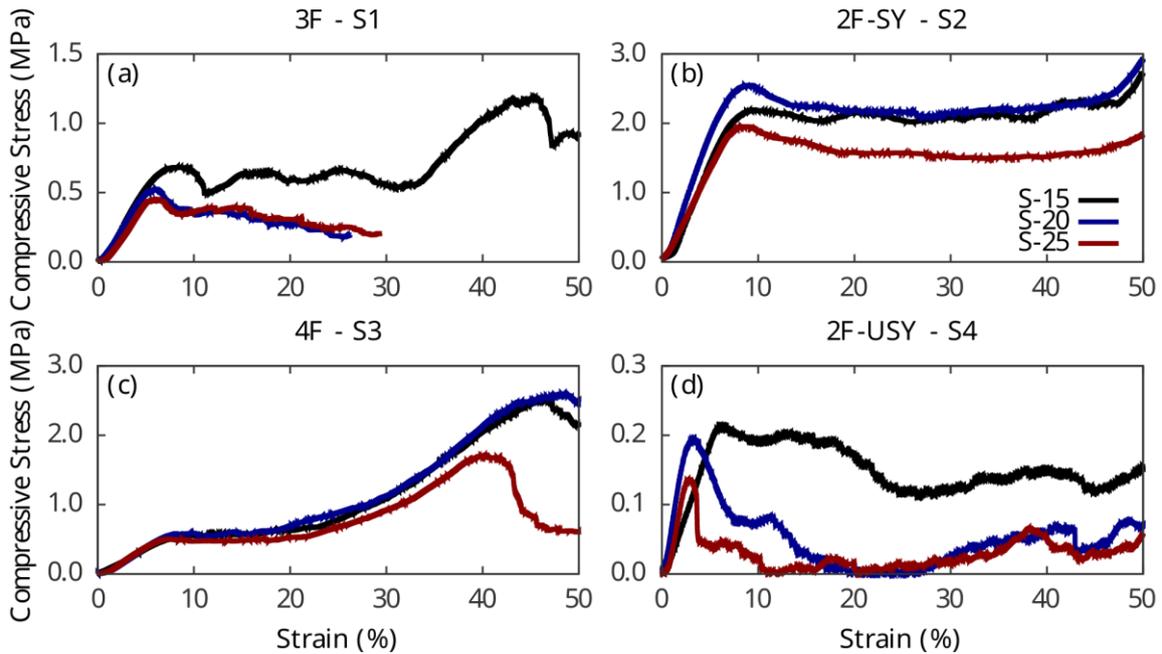

**Fig. 4.** Experimental results from the compression test. Compressive Stress (MPa) versus Strain (%) curves for the four structures: (a) 3F - S1, (b) 2F-SY - S2, (c) 4F - S3, and (d) 2F-USY - S4. The lines compare the effect of different sample dimensions: S-15 (black), S-20 (blue), and S-25 (red).

The 2F-SY structure consistently exhibits the highest stiffness and peak stress among all specimens (see Fig. 4(b)). Its symmetric sublattice arrangement leads to uniform load transfer throughout the network, which delays the onset of local structural instabilities. As a consequence, the 2F-SY curves show a well-defined elastic region and a stable plateau (plastic-like regime) extending up to ~40% strain, indicating efficient stress redistribution and ductile-like compaction behavior. Our protocol of increasing the sample size from 15 to 25 mm resulted in a small reduction in peak stress, consistent with the typical effects of geometric scaling and printing-induced imperfections.

We can se that the 3F architecture also demonstrates a stable response. The curves exhibit mild oscillations in the plateau region (see Fig. 4(a)). The presence of three intertwined sublattices enhances structural redundancy. This enables the structure to still sustain load after the collapse of individual struts. Although less pronounced than in 2F-SY, the 3F specimens exhibit smooth compaction and gradual energy absorption. This effect is particularly noticeable in the S-20 and S-25 samples.

The 4F structure (Fig. 4(c)) exhibits an intermediate behavior between 2F-SY and 3F. It exhibits relatively high stiffness and pronounced fluctuations in the plateau region. In this case, plateau fluctuation refers to a force-displacement curve that exhibits noticeable rises and drops rather than a smooth, flat response. This behavior is due to its higher internal connectivity, which makes it more rigid at the beginning and also creates multiple pathways for structural failures. As the strain increases, the 4F lattices form localized buckling zones that spread diagonally, creating a serrated plateau common in cellular materials constrained in multiple directions.

On the other hand, the 2F-USY structure has the lowest stiffness and strength at all sizes (see Fig. 4(d)). The uneven arrangement of its sublattice causes stress concentrations and makes it structurally unstable at the first stages of compression, as evidenced by the sharp oscillations and early peak in Fig. 4(d). Diagonal shear is the dominant deformation mode, and densification begins at much lower strain levels than in the other architectures. The scaling effect is also more substantial: larger samples (S-25) show a clear drop in their ability to bear weight, probably because the imperfections along the weak direction of the asymmetric network are more pronounced.

The experimental results show that the sublattice topology has a significant effect on the mechanical performance of Diamondiyne-derived architectures. Symmetric or higher-order interlocked networks (2F-SY, 3F, 4F) facilitate uniform deformation, smoother plateau regions, and enhanced energy absorption, whereas asymmetric architectures (2F-USY) exhibit increased sensitivity to instability and geometric imperfections. Although the sizes differ, the qualitative behavior of each topology remains the same. This indicates that geometry is the primary factor affecting the compressive response.

It is essential to consider specimen size and the topology's apparent effect on the compressive response. We tested each Diamondiyne-derived architecture at three sizes (15, 20, and 25 mm) to assess how size affects stiffness, strength, and overall deformation behavior. Therefore, one can conclude that smaller specimens consistently exhibit enhanced mechanical performance, characterized by higher peak stress, a longer plateau region, and delayed densification.

Geometric scaling primarily controls this behavior. As specimen size increases, strut length increases, and relative density decreases. This trend makes the lattice more porous and less efficient structurally. Longer struts are more likely to bend and become structurally unstable prematurely, reducing their overall stiffness and strength. The 15 mm samples, on the other hand, have a denser and more geometrically complex network, which helps transfer loads and delays failure. These observations align with the established scaling laws for bending-dominated cellular materials, which indicate that mechanical properties diminish as relative density decreases [11] [28].

**Table 2** presents the main mechanical features derived from the experimental stress-strain curves. It also gives quantitative proof that the mechanical performance depends heavily on the underlying sublattice topology. The 2F-SY (S2) architecture consistently outperforms on most features, while the 2F-USY (S4) architecture consistently performs the worst. The 3F (S1) and 4F (S3) structures are in different but similar mechanical regimes. This shows how symmetry and the number of interpenetrating frameworks affect the efficiency of load transfer.

**Table 2.** Specific yield strength, specific resilience, specific energy absorption, yield load, and density obtained from the experimental compression tests for all Diamondiyne-derived architectures (S1–S4) at three nominal sizes (15, 20, and 25 mm).

|   |   | Specific Yield strength (MPa·g$^{-1}$·cm³) | Specific Resilience (J·g$^{-1}$) | Specific Energy consumption (J·g$^{-1}$) | Load$_{Yield}$ (N) | Density (g·cm$^{-3}$) |
|---|---|---|---|---|---|---|
| S1 | 15 | 2.06 ± 0.01 | 6.70 ± 0.15 | 106.09 ± 4.88 | 134.62 ± 1.21 | 0.290 ± 0.009 |
| S1 | 20 | 1.77 ± 0.05 | 4.21 ± 0.34 | 32.15 ± 1.88 | 189.39 ± 5.85 | 0.268 ± 0.002 |
| S1 | 25 | 1.53 ± 0.10 | 3.43 ± 0.39 | 29.67 ± 3.27 | 254.64 ± 16.68 | 0.266 ± 0.002 |
| S2 | 15 | 4.91 ± 0.08 | 19.09 ± 1.02 | 264.35 ± 2.53 | 700.00 ± 11.92 | 0.377 ± 0.003 |
| S2 | 20 | 5.91 ± 0.20 | 20.41 ± 0.98 | 279.01 ± 3.2 | 1495.75 ± 25.00 | 0.378 ± 0.004 |
| S2 | 25 | 5.63 ± 0.08 | 18.86 ± 0.69 | 245.09 ± 6.16 | 1853.72 ± 26.60 | 0.313 ± 0.002 |
| S3 | 15 | 1.58 ± 0.06 | 4.73 ± 0.23 | 247.26 ± 1.73 | 109.04 ± 3.90 | 0.307 ± 0.007 |
| S3 | 20 | 1.71 ± 0.09 | 5.21 ± 0.17 | 254.52 ± 8.5 | 206.18 ± 11.17 | 0.302 ± 0.012 |
| S3 | 25 | 1.62 ± 0.07 | 5.27 ± 0.07 | 158.07 ± 5.97 | 296.59 ± 12.69 | 0.293 ± 0.013 |
| S4 | 15 | 1.03 ± 0.03 | 2.17 ± 0.66 | 54.15 ± 4.03 | 60.83 ± 2.02 | 0.197 ± 0.003 |
| S4 | 20 | 0.99 ± 0.02 | 1.48 ± 0.19 | 23.11 ± 2.63 | 91.89 ± 1.61 | 0.186 ± 0.002 |
| S4 | 25 | 0.79 ± 0.06 | 0.77 ± 0.06 | 16.13 ± 0.98 | 102.70 ± 7.67 | 0.156 ± 0.007 |

The S-20 case has a specific yield strength of 5.91 MPa·g$^{-1}$·cm³, approximately three and six times that of the 3F S-20 and 2F-USY S-20, respectively, indicating that 2F-SY is the strongest structure. The symmetric sublattice arrangement of this material improves the performance by distributing stress more uniformly and delaying the onset of buckling instabilities. The 2F-USY architecture, by contrast, exhibits the lowest strength across all sizes, with values as low as 0.77 MPa·g$^{-1}$·cm³. This is consistent with the early shear-dominated structural collapse seen in its stress–strain response.

Specific resilience and specific energy absorption exhibit similar trends. The 2F-SY structure has a maximum specific resilience of 20.41 J·g$^{-1}$ and a specific energy absorption of 279.01 J·g$^{-1}$, indicating its ability to store/release mechanical energy during compression. The 3F and 4F architectures can also absorb a substantial amount of energy, up to 264.35 J·g$^{-1}$ and 254.52 J·g$^{-1}$, respectively. However, their performance is more affected by geometric scaling, especially in the larger S-25 specimens. This behavior indicates that it is more difficult to redistribute loads in networks with higher sublattice multiplicity because competing collapse pathways lead to greater fluctuations in deformations.

The 2F-USY architecture, by contrast, exhibits a significant drop in energy-related metrics, particularly for larger sample sizes. The S-25 sample requires only 16.13 J·g$^{-1}$, which is nearly 10 times lower than that of the other architectures. This result confirms that asymmetric sublattice alignment creates stress concentrations that are detrimental to the structure's performance, reduces redundancy, and reduces stability under compression. The load data can provide further information about how the system scale can affect the results. Larger samples tend to carry more weight. However, their normalized performance (per mass or volume) decreases, particularly for 4F and 2F-USY. This occurs because printing causes imperfections to accumulate over larger areas and because long, unsupported strut lengths lead to buckling at shorter lengths.

Although the density values are very similar across all architectures, they do show small but significant differences. The 2F-SY and 3F structures have slightly higher densities because their internal connections are more evenly spread out. The 2F-USY, on the other hand, is lighter, but its irregular arrangement makes it mechanically weaker, as the structure lacks a balanced stress distribution. These changes in density support the conclusion that 2F-SY's superior performance is attributable to its topological properties rather than its mass. **Figure 5** shows how specimen size affects four key mechanical metrics: load yield, specific yield strength, specific resilience, and specific energy consumption. A consistent trend, which depends on the topology, appears across all panels: smaller specimens perform better, and the ranking of architectures is the same as that observed in the stress-strain curves.

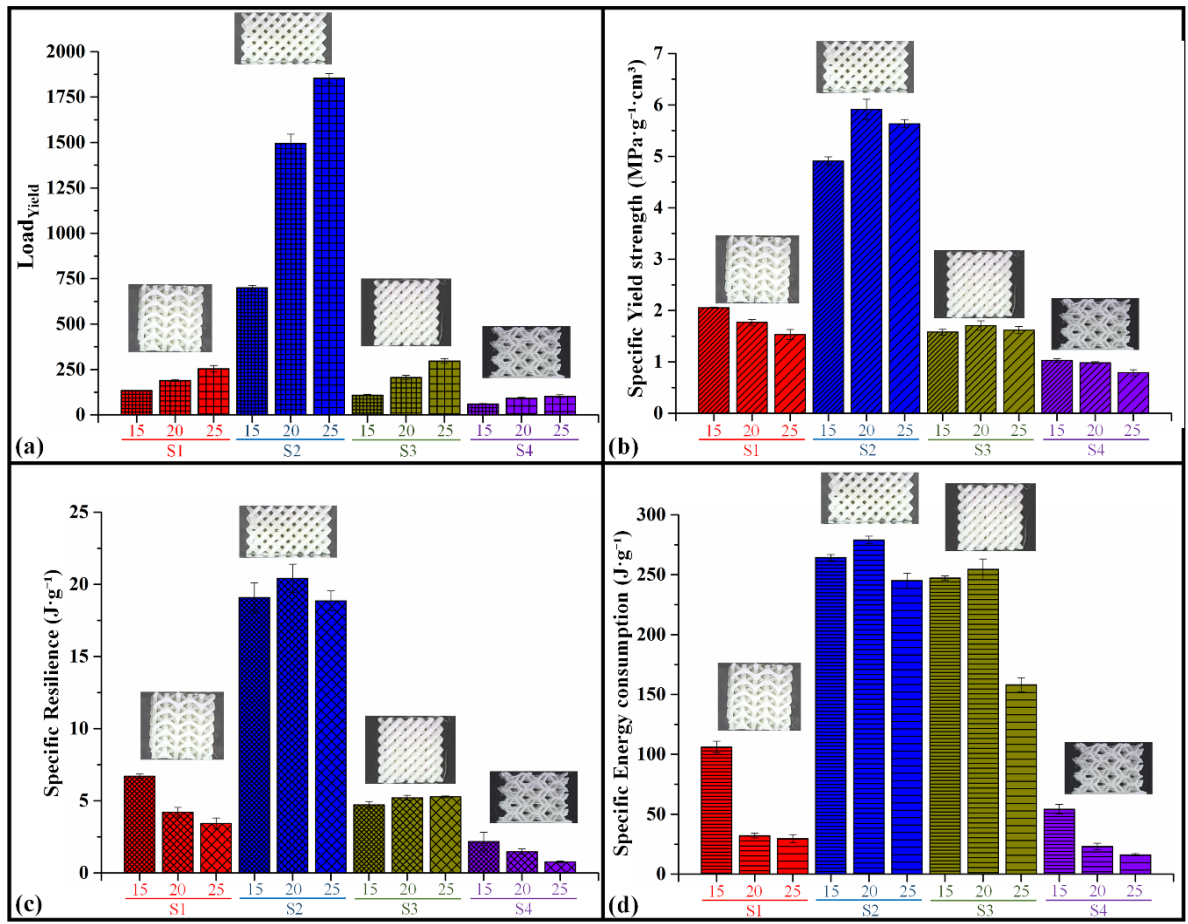

**Fig. 5.** Effect of specimen dimension on key mechanical properties of the four Diamondiyne-derived architectures. (a) Load$_{yield}$, (b) specific yield strength, (c) specific resilience, and (d) specific energy consumption for structures S1 (3F), S2 (2F-SY), S3 (4F), and S4 (2F-USY) at three nominal sizes (15, 20, and 25 mm). Error bars represent the standard deviation across three independent tests.

**Fig. 5(a)** shows that 2F-SY (S2) reaches the highest Load$_{yield}$, with values exceeding 1800 N for the S2–25 specimen and 1500 N for S2–20. This ordering reflects the strong and uniform load-transfer pathways created by its symmetric two-sublattice configuration. In contrast, 2F-USY (S4) has the lowest Load$_{yield}$ for all dimensions, which is consistent with the early buckling and diagonal shear failure observed during testing. Structures 3F (S1) and 4F (S3) present intermediate regimes, although 4F produces slightly higher loads due to its greater connectivity and redundancy of struts. A similar hierarchy for specific yield strength is revealed in **Fig. 5(b)**, which normalizes the response by specimen mass. The S2 architecture once again dominates, reaching 5.9 MPa·g$^{-1}$·cm³, followed by S3 and S1. The S4 structure shows the lowest specific yield strength, dropping below 1.0 MPa·g$^{-1}$·cm³ for the larger specimens. These results confirm that topology, not mass, is the primary determinant of yield behavior: symmetric or higher-order interpenetrated networks resist instability more efficiently, whereas asymmetric frameworks like 2F-USY concentrate stress along weaker directions. **Fig. 5(c)** examines specific resilience, which represents the elastic energy stored before yielding. The S2 architecture, one more time, exhibits the highest values, exceeding 20 J·g$^{-1}$, indicating efficient elastic deformation before plateau collapse. The S3 and S1 networks exhibit moderate resilience, while S4 exhibits minimal energy storage (≈1–3 J·g⁻¹), further reflecting its susceptibility to early instability. In **Fig. 5(d)**, the specific energy consumption is compared, capturing the total energy absorbed before densification. The S2 and S3 structures show the highest performance (≈245–280 J·g⁻¹), with S1 showing intermediate behavior and S4 absorbing the least energy (≈16–23 J·g⁻¹). This ordering is fully consistent with the deformation modes observed experimentally: S2 and S3 undergo stable and progressive compaction, whereas S4 collapses along a preferred shear plane, limiting its energy absorption capacity.

### 3.2. Directional compressive response from molecular dynamics

We now analyze the compressive stress–strain curves obtained from reactive molecular dynamics simulations, considering the four Diamondiyne-derived architectures under loading along the three principal directions (see **Fig. 6**). In all cases, the curves display the typical features of bending-dominated architected networks: an initially linear regime, followed by progressive stiffening and, eventually, a rapid stress increase associated with the collapse of the internal porosity. Absolute stress levels are in the gigapascal range, as expected for carbon frameworks, and therefore cannot be directly compared to the experimental PLA results. Nevertheless, the relative trends and rankings among architectures provide valuable insights into the roles of topology and sublattice arrangement, and can be meaningfully contrasted with the experiments.

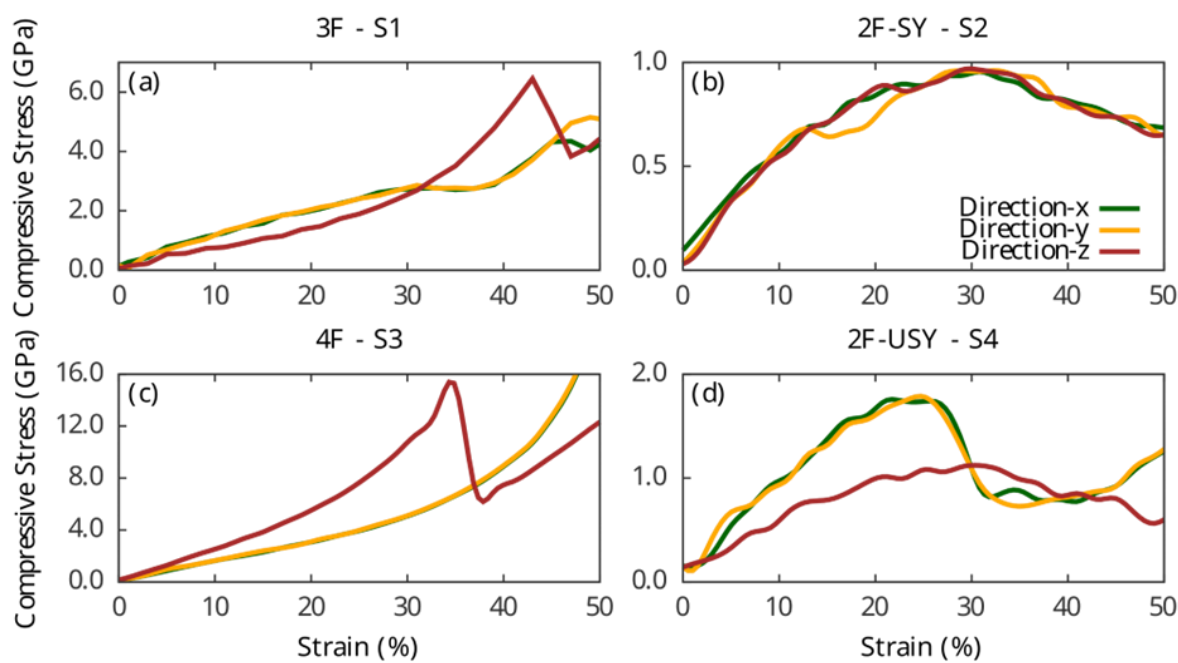

**Fig. 6.** Directional compressive response simulated via molecular dynamics using the ReaxFF force field. Compressive Stress (GPa) versus Strain (%) curves for the four structures: (a) 3F - S1, (b) 2F-SY - S2, (c) 4F - S3, and (d) 2F-USY - S4. The lines compare the mechanical behavior under compression along the three principal directions: the x-direction (green), the y-direction (yellow), and the z-direction (red).

For the 3F architecture (S1, **Fig. 6(a)**), the three directions exhibit similar responses up to ≈30% strain, indicating an almost isotropic behavior at small deformations. At higher strains, loading along the z-direction becomes slightly stiffer than along x and y, which is consistent with the higher connectivity of load paths in that direction. Experimentally, the 3F specimens also showed an intermediate stiffness and a relatively smooth compaction, aligning with the simulation result that this architecture distributes load comparatively uniformly across directions.

The 2F-SY structure (S2, **Fig. 6(b)**) shows only modest directional differences throughout the entire strain range, reflecting the symmetric arrangement of its two interlocked sublattices. This quasi-isotropic response from MD is entirely consistent with the experiments, where 2F-SY exhibited the highest and most stable mechanical performance among all printed architectures, with well-defined plateaus and delayed densification. Both scales, therefore, indicate that sublattice symmetry promotes uniform load transfer and reduces directional sensitivity.

In contrast, the 4F architecture (S3, **Fig. 6(c)**) displays pronounced anisotropy. The curve for the z direction is stiffer and exhibits the highest stresses, while the curves for the x and y directions are smoother and exhibit a more gradual stiffening. This trend suggests that the four interpenetrating sub-networks create preferential load paths along z, which can be exploited to adjust stiffness directionally. In experiments, the 4F samples also ranked among the stiffest structures, although 2F-SY slightly outperformed them in specific yield strength and energy absorption. This difference is reasonable: the simulations consider a defect-free, fully periodic carbon lattice, whereas the experiments involve printed PLA with surface roughness, interlayer adhesion effects, and slight density variations that may penalize the more complex 4F geometry.

The 2F-USY architecture (S4, **Fig. 6(d)**) exhibits the most substantial directional dependence. The load along the z-axis stiffens the response and increases stress levels, whereas the loads along the x- and y-axes make the response much less stiff and soften it sooner. The strong anisotropy arises from the fact that the two sub-networks are arranged unevenly, resulting in weak directions with fewer practical load-bearing elements. The experimental results qualitatively corroborate this trend: 2F-USY consistently exhibited the lowest stiffness, yield strength, and energy absorption, and also showed diagonal shearing and early damage localization. Thus, both MD and experiments identify 2F-USY as the mechanically least efficient architecture, particularly when loaded along unfavorable directions.

**Table 3** summarizes the directional Young's moduli obtained from the initial linear portion of the stress-strain curves of the molecular dynamics. The results show that the stiffness of Diamondiyne-derived structures depends primarily on their subnetwork pattern, with each structure exhibiting a unique directional stiffness. Although the actual numbers differ from the measured stiffness due to differences in material (carbon vs. PLA), strain rate, and edge effects, the order and general trends remain the same across different configurations and/or scales.

**Table 3.** Directional Young's modulus obtained from molecular dynamics simulations for the four Diamondiyne-derived architectures under uniaxial compression along the x, y, and z directions.

| Structure | Young's modulus (GPa) | | |
| :---: | :---: | :---: | :---: |
| | Direction x | Direction y | Direction z |
| 2F-SY | 5.93 | 5.82 | 5.87 |
| 2F-USY | 7.62 | 9.99 | 4.74 |
| 3F | 14.56 | 14.14 | 7.15 |
| 4F | 15.90 | 15.57 | 24.13 |

The 4F architecture is the stiffest of all structures, particularly along the z-direction, where it has a value of 24.13 GPa, nearly twice as stiff as along the x- and y-directions. This strong anisotropy arises from four interpenetrating sublattices that preferentially direct load along the z-axis. The 4F specimens were also among the stiffest lattices in the experiments, exhibiting high yield loads and among the largest specific energy absorptions. This is in line with a framework that redistributes load along well-connected axes. The 3F structure also exhibits moderate stiffness in all directions (~14 GPa in x and y), but is slightly less stiff along z. This trend is similar to that observed in the experiments, in which 3F specimens exhibited intermediate yield strength and stable compaction. This suggests a load-carrying network that is balanced but has some directional dependence.

The 2F-SY structure, on the other hand, has nearly identical directional moduli (5.82–5.93 GPa) along the axes, indicating a quasi-isotropic, symmetric two-sublattice structure. This isotropy is similar to the conclusions of experiments that demonstrated that 2F-SY exhibited the most consistent and predictable mechanical response across all sample sizes, as well as the highest specific resistance to creep and energy absorption. The structure of 2F-USY is different because it exhibits high anisotropy, with stiffness ranging from 4.74 GPa (along the z direction) to 9.99 GPa (along the y direction). The fact that its interconnected subnetworks are not symmetrical makes the load transfer pathways irregular and the directions weaker. In experiments, this manifests as early buckling, diagonal shear failure, and the lowest yield strength and energy absorption among all architectures. The relationship between the anisotropy revealed by MD and the mechanical instability seen in the printed samples supports the idea that subnetwork asymmetry greatly weakens structural integrity.

### 3.3. Deformation mechanisms: experimental and atomistic insights

**Fig. 7** illustrates the 3F architecture deforming under uniaxial compression, accompanied by MD simulations at matching strain levels. Both results show the 3F structure deforms progressively and stably through distributed buckling and gradual compaction of the angled struts.

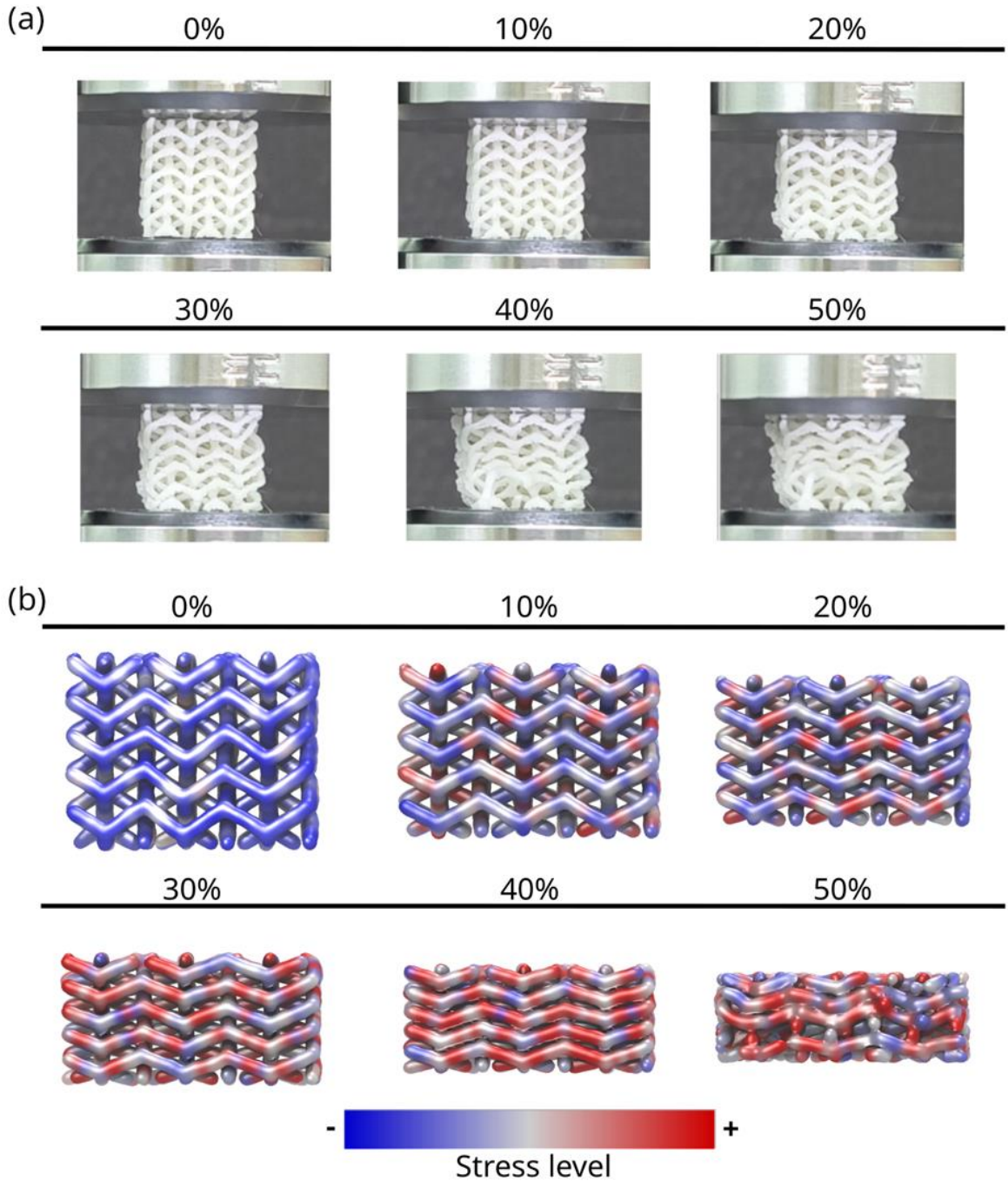

**Fig. 7.** (a) Experimental images of the uniaxial compression test of the 3F structure at different strain levels (0–50%), showing the progressive collapse and rearrangement of the lattice during loading. (b) Corresponding computational simulations of the 3F structure under the same strain levels, highlighting the stress distribution along the structural links, where blue regions indicate low-stress areas and red regions correspond to high-stress concentrations.

In the experimental images, the onset of deformation is observed at approximately 10–20%, when the upper layers of the structure begin to bend evenly (see **Fig. 7(a)**). The structure does not exhibit localized instability; instead, it has a uniform deformation field throughout. When deformation reaches 30–40%, the diagonally oriented struts bend inward in a coordinated manner, while the overall shape remains unchanged. At 50% deformation, the structure is highly compacted, yet it retains a recognizable network structure. This illustrates how the three-subnetwork configuration operates mechanistically. The experimental stress-strain curve for the 3F sample exhibits a smooth plateau region with small oscillations and delayed densification relative to other architectures, such as 2F-USY. This behavior is in line with that.

The MD simulations depicted in **Fig. 7(b)** replicate these deformation characteristics at the nanoscale. When the strain is low (0–10%), the stress remains low and relatively uniform across the lattice. This trend indicates the rigidity of the interconnected carbon structure in isolation (non-interacting). Stress accumulates along the same diagonal and vertical members that bend in experiments as deformation increases to 20–30%. This trend indicates that the simulated structure depicts the primary mechanisms by which the load is carried, accumulated, and released. For higher deformation regimes (40–50%), MD stress maps indicate that stress tends to be accumulated substantially in regions that bend during testing. This behavior suggests that the 3F structure absorbs energy by deforming discretely rather than collapsing all at once.

**Fig. 8** compares the experimental compression sequence of the symmetric two-sublattice architecture 2F-SY with corresponding molecular dynamics simulations at equivalent strain levels from 0% to 50%. These results provide a multiscale view of the stable and uniform deformation that characterizes the 2F-SY lattice. Experimentally, this lattice achieves the highest strength and energy absorption among all Diamondiyne-derived structures (see **Fig. 8(a)**). The experimental images reveal that deformation begins gradually, with the lattice maintaining its overall geometric topology up to approximately 20% strain. At this point, the struts are beginning to bend slightly, and the cell openings are becoming smaller evenly along the height of the sample. The 2F-SY architecture does not exhibit early shear bands or localized collapse structures, unlike asymmetric arrangements such as 2F-USY. Instead, the deformation is spread evenly, indicating that the sub-networks' symmetric arrangement facilitates even load transfer and delays the onset of structural instability. As deformation reaches 30–40%, the network becomes denser, maintaining the structure's stability. The sample shows an almost ordered compaction up to 50% of deformations without complete structural failure. This behavior is fully consistent with the experimental stress–strain curves, where 2F-SY exhibits the highest plateau stress, the most stable post-yield region, and the largest specific energy absorption.

The MD snapshots shown in **Fig. 8(b)** reproduce this smooth and uniform deformation pattern at the atomistic scale. At small strain levels (0–10%), the stress distribution spreads across the whole structure, with only a few small peaks. When the strain exceeds 20%, the calculated stress maps indicate that moderate stress concentrations begin to form in the diagonal struts. However, these areas progress slowly rather than suddenly. The simulations show that stress accumulates throughout the lattice at 40–50% strain, but the overall structure remains unchanged. This progression is similar to that observed when the material was compacted in the laboratory, indicating that the primary mechanism of deformation is distributed bending and sequential buckling of the symmetric strut network.

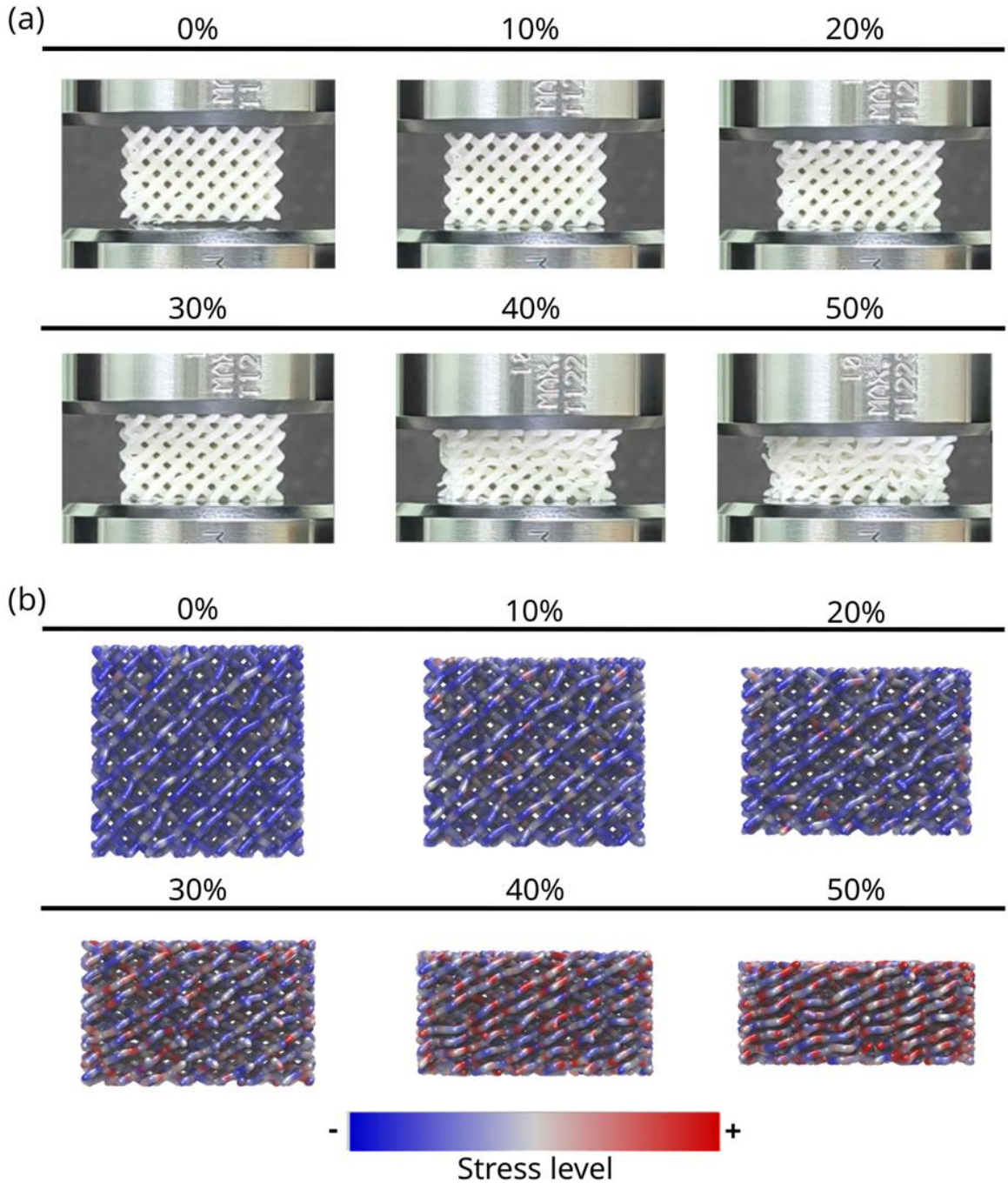

**Fig. 8.** (a) Experimental uniaxial compression sequence of the 2F-SY structure at different strain levels (0–50%), showing the gradual deformation and densification of the lattice under loading. (b) Corresponding molecular dynamics simulations of the same structure under equivalent strain conditions, illustrating the evolution of local stress distribution throughout the deformation process. The color map represents the stress level, with blue and red corresponding to low- and high-stress regions, respectively.

A key observation from both experiment and simulation is the high degree of isotropy in the compressive response of 2F-SY. The symmetric placement of the two interpenetrated sublattices yields nearly identical stiffness along the principal directions in MD, and this structural isotropy manifests macroscopically as a uniform collapse free of directional bias. This differs from the 2F-USY architecture, in which asymmetry leads to early diagonal shear, directional dependence, and reduced load capacity. The consistency between scales confirms that the superior mechanical performance of the 2F-SY experimentally, its high specific yield strength, resilience, and energy absorption, directly reflects the inherent stability provided by its symmetric topological design.

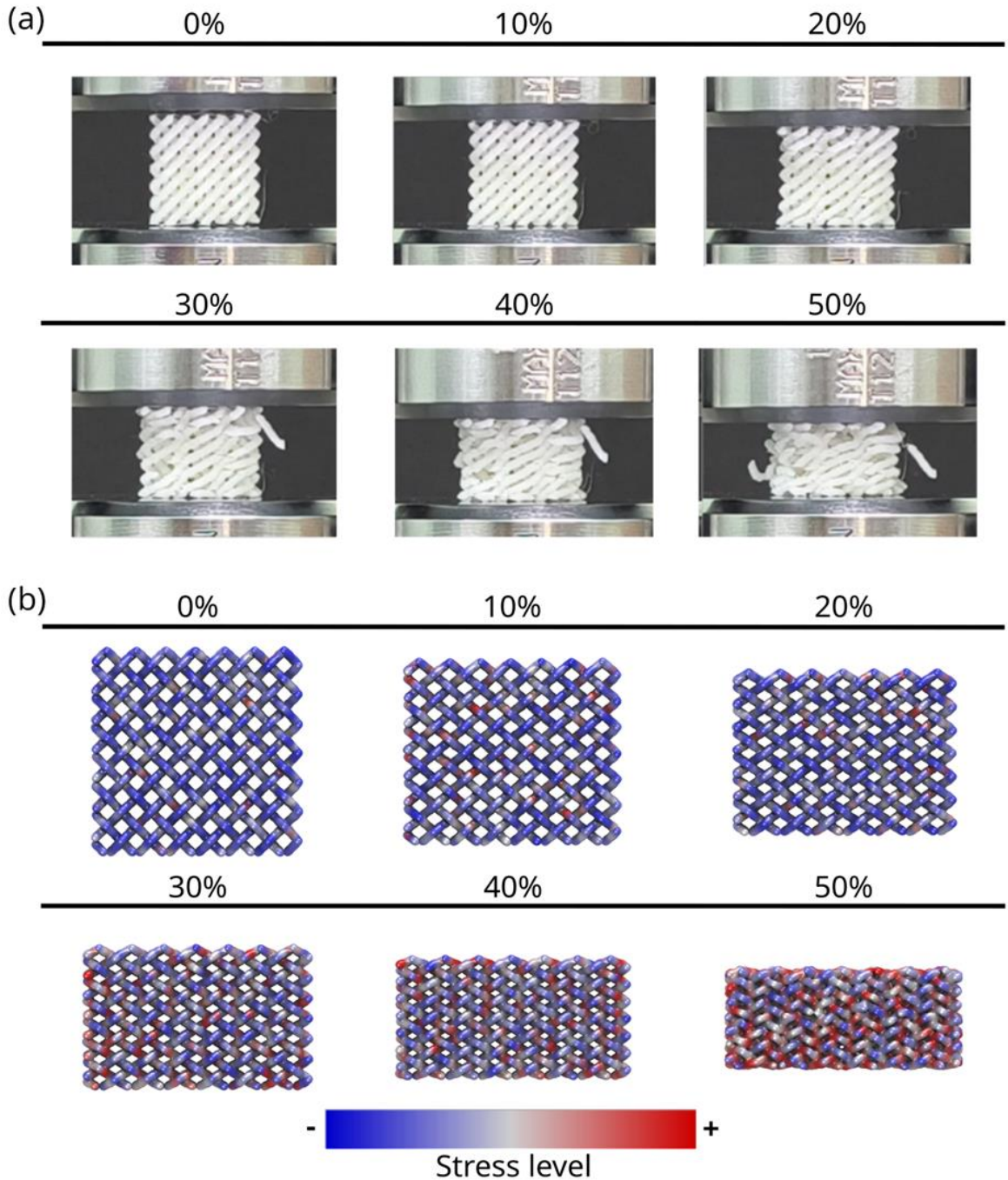

**Fig. 9.** (a) Experimental snapshots of the 4F structure subjected to uniaxial compression at strain levels from 0% to 50%, exhibiting enhanced structural interlocking and stability during deformation. (b) Simulated stress maps of the 4F structure under identical strain conditions, evidencing a homogeneous stress distribution at low strains and pronounced stress concentration at higher compression levels. The stress intensity scale ranges from low (blue) to high (red).

**Fig. 9** presents the deformation sequence of the 4F architecture under uniaxial compression. It also displays molecular dynamics simulations that illustrate the evolution of local stress at equivalent strain levels. This structure is composed of four interpenetrated sublattices and is the most topologically complex among the architectures examined. The multiscale results reveal how this complexity influences its deformation behavior. The experimental snapshots in **Fig. 9(a)** demonstrate that the 4F structure preserves a highly ordered lattice configuration during the initial phases of compression. Up to approximately 20% strain, the deformation is predominantly elastic, and the overall shape remains the same. When the strain reaches 30%-40%, the angled struts begin to bend inward. The collapse remains largely homogeneous across the specimen height. At a 50% strain, the lattice undergoes significant densification while retaining a coherent structural pattern, without exhibiting abrupt failure or shear localization. This behavior is consistent with the experimental stress–strain response of the 4F specimens. They exhibit intermediate-to-high stiffness and a stable plateau region, particularly compared with the asymmetric 2F-USY architecture.

MD simulations shown in **Fig. 9(b)** reveal a similar deformation pattern at the nanoscale. At small strain levels (0–10%), the stress distribution is nearly uniform. This reflects the highly redundant load-bearing network formed by the four interpenetrated frameworks. As deformation rises to 20-30%, localized stress concentrations emerge along the oriented struts in the diagonal. But these areas are still widely distributed and do not cause things to early break. Simulations show that the stress intensity along the vertical and diagonal struts increases substantially when the deformation regime is higher (40–50%). Even under this regime, the distribution remains spatially widespread. This distributed stress accumulation correlates with the progressive experimental densification. It shows that the 4F architecture dissipates energy by bending and buckling at multiple locations rather than locally structurally failing. The way 4F changes shape is related to its structure. Four interpenetrating sublattices create multiple parallel load paths. This makes the structure more redundant. This allows the lattice to stand higher load before becoming denser. This is consistent with both the MD-derived Young's modulus, particularly the very high stiffness along the z-direction, and the experimental mechanical features. The 4F shows one of the highest specific energy consumption values, surpassed only by the symmetric 2F-SY. Importantly, the multiscale results highlight that although the 4F lattice experiences higher localized stresses than the 3F structure, these stresses do not translate into catastrophic structural failure. This is due to the multiple pathways to release stress of the load-carrying struts.

Finally, **Fig. 10** compares the compressive deformation of the asymmetric two-sublattice architecture (2F-USY) observed experimentally together with the corresponding molecular dynamics simulations at matched strain levels. Among all Diamondiyne-derived structures, 2F-USY exhibits the most unstable mechanical behavior. The multiscale results in this figure reveal how the asymmetric arrangement of its sublattices fundamentally undermines its ability to redistribute load.

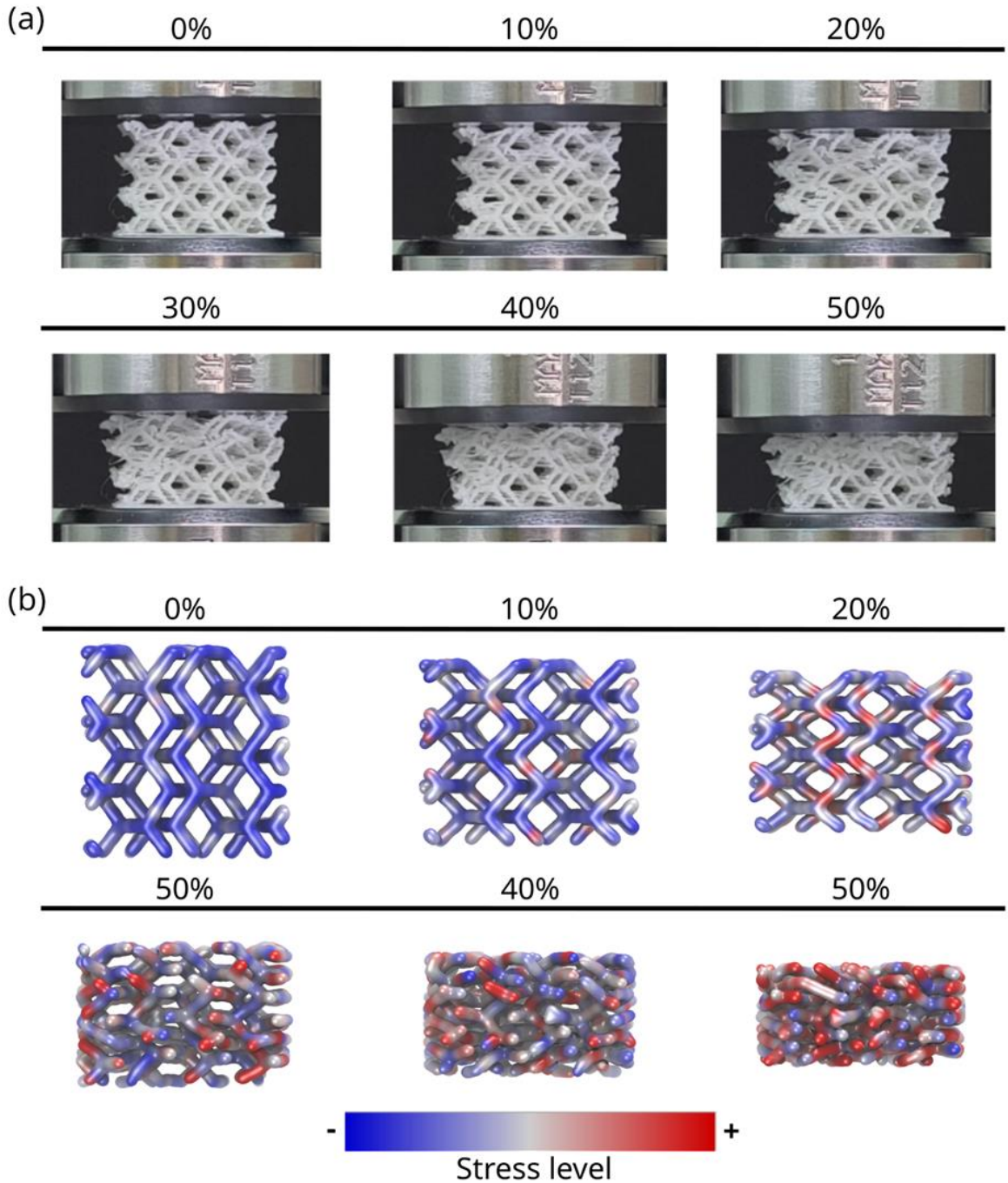

**Fig. 10.** (a) Experimental compression images of the 2F-USY structure at increasing strain levels (0–50%), revealing a distinct deformation pattern compared to the symmetric 2F-SY configuration. (b) Computational simulations of the 2F-USY structure under the same strain range, highlighting asymmetric stress localization and progressive structural densification. Blue and red colors denote regions of lower and higher stress concentration, respectively.

The experimental images in **Fig. 10(a)** show that deformation starts earlier and progresses less uniformly in 2F-USY compared with the other architectures. Even at a moderate strain level (≈10%), the lattice exhibits a visible tilting and uneven shortening of the top layers, indicating the presence of preferred weak directions within the structure. By a strain of 20–30%, the deformation becomes strongly localized, forming a diagonal shear band that propagates across the lattice height. This collapse, which is primarily due to shear, is very different from the progressive and uniform densification seen in 3F, 4F, and especially 2F-SY. When the strain level is higher (40–50%), the 2F-USY structure loses its structural integrity, and the collapse happens in a small area instead of spreading out across the whole lattice. This localized instability explains the markedly lower plateau stress, reduced energy absorption, and premature densification observed in the experimental stress–strain curves.

The MD simulations in **Fig. 10(b)** are in strong qualitative agreement with these experimental observations. At a small strain level, the stress distribution already exhibits significant anisotropy, with stress concentrating along the struts aligned with the weaker sublattice direction. As strain rises to 20–30%, these stress concentrations become more pronounced and start to align along diagonal paths, mimicking the early onset of shear instabilities observed in experiments. At 40–50% strain, the nanoscale stress field exhibits a transparent gradient aligned with the collapse direction. This shows that the asymmetric layout cannot evenly distribute the load across the structure. The MD results indicate that the asymmetry between the two interpenetrating sublattices leads to an imbalance in structural redundancy. This makes some directions fundamentally more rigid and others more prone to srtructural instability.

**3.4 Performance comparison with other structures**

The performance of the present Diamondiyne structures was compared with the previous reported structures made of PLA, as shown in **Fig. 11**. Density, specific energy consumption, specific resilience and specific yield strength were considered as standard parameters for different structures. Having a lower-density structure can improve performance, reduce material consumption, and result in a lightweight structure. In terms of density, the present structure has advantages (less than 0.3 g·cm$^{-3}$) compared to Schwarzites, zeolite-type structures [29] [22] (more than 1 g·cm$^{-3}$), and Schwarzite and Schwarzyne-based structures (0.5-0.7 g·cm$^{-3}$) [20]. This may be due to a unique feature of movable interlocked lattices. In terms of specific energy consumption, the present two diamondiyne structures (S1-15 and S4-15) exhibited equivalent performance compared to the previous Schwarzites, zeolite-type structures (50-100 J·g$^{-1}$), whereas S3-20 and S4-20 performed exceptionally well, having a value of 250-280 J·g$^{-1}$, despite a lower structure density. The present structure has given an average performance in terms of specific yield strength compared to a few Schwarzite and Zeolite [22] structures, which varied from 0 to 5 MPa·g$^{-1}$·cm³; also, it has given lower values compared to a few Schwarzite [6], Zeolite [22], and Schwarzite-Schwarzyne-based structures [20]. Among the present diamondiyne structures, 2F-SY with a dimension of 20mm (S2-20) has performed exceptionally well with a density, specific resilience, and specific energy consumption of 0.38 g·cm$^{-3}$, 20.41 J·g$^{-1}$, and 279.01 J·g$^{-1}$, respectively.

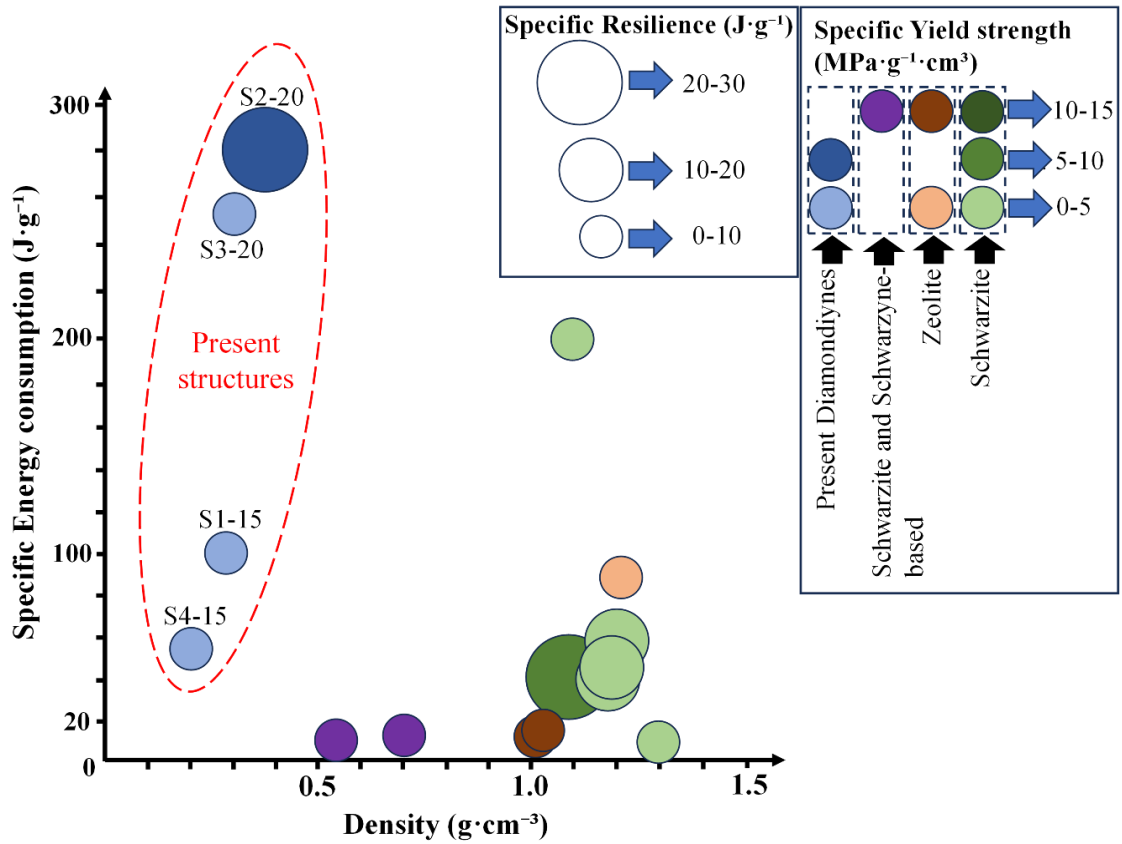

**Fig. 11**. Performance map of present diamondiyne compared with Schwarzites and zeolite-type structures.

## 4. Conclusions

In summary, we conducted a multiscale mechanical study of Diamondiyne-derived materials combining uniaxial compression tests on 3D-printed specimens with reactive MD simulations of corresponding atomic-scale models. Four different topological configurations were considered: 3F, 2F-SY, 4F, and 2F-USY. The main goal was to investigate how sublattice symmetry, multiplicity, and interpenetration affect the structural response across different length scales. The experimental findings indicated that the geometric configuration primarily influences the compressive behavior of these lattices, whereas specimen dimensions have only a marginal effect.

The symmetric two-sublattice structure (2F-SY) exhibited the highest specific yield strength, resilience, and energy absorption among the architectures. The asymmetric 2F-USY configuration, by contrast, performed poorly due to early localized structural instabilities. The 3F and 4F architectures exhibited similar mechanical behavior, but the 4F structure has more redundant load paths and greater directional stiffness. MD simulations qualitatively confirmed these trends by replicating the deformation mechanisms observed experimentally and demonstrating nanoscale stress redistribution patterns that directly correspond to the underlying lattice topology.

The current structures have a density of 0.3 g·cm$^{-3}$, which is much lower than that of Schwarzites, zeolites, and structures based on Schwarzite and Schwarzyne. This feature improves performance and reduces material and weight.

This study shows that the mechanical strength of lattices made from Diamondiyne is primarily determined by the arrangement of their sublattices, rather than by the materials from which they are made or their size. The close agreement between the experiment and the simulation indicates that movable and interlocked carbon frameworks could be a promising approach for designing lightweight, energy-absorbing metamaterials. These results create new opportunities for engineering architected materials whose macroscopic performance can be customized via atomic-scale topological design.

**Acknowledgments:** The authors gratefully acknowledge financial support from the following institutions: the National Council for Scientific and Technological Development (CNPq), the Federal District Research Support Foundation (FAPDF), and the Coordination for the Improvement of Higher Education Personnel (CAPES), and PETRONAS Brazil LTDA. The authors also express their gratitude to the National Laboratory for Scientific Computing for providing resources through the Santos Dumont supercomputer and to the Centro Nacional de Processamento de Alto Desempenho em São Paulo (CENAPAD-SP) for support related to project 634. Additional resources were provided by Lobo Carneiro HPC (NACAD) at the Federal University of Rio de Janeiro (UFRJ). L.A.R.J. acknowledges financial support from FAPDF grant 0193.000942/2015, CNPq grants 307345/2021-1 and 350176/2022-1, FAPDF-PRONEM grant 00193.00001247/2021-20, and PDPG-FAPDF-CAPES Centro-Oeste grant number 00193-00000867/2024-94. C.S.T. acknowledges Core research grant of SERB, India, STARS projects by MHRD-India, DAE Young Scientist Research Award (DAEYSRA), and AOARD (Asian Office of Aerospace Research and Development) grant no. FA2386-21-1-4014, and Naval research board for funding support. D. S. Galvao also acknowledges INEO/CNPq and FAPESP Grant **2025/27044-5**.

**Conflicts of Interest:** There are no conflicts of interest to declare.

**References**

[1] H.W. Kroto, J.R. Heath, S.C. O'Brien, R.F. Curl, R.E. Smalley, C60: Buckminsterfullerene, Nature 1985 318:6042 318 (1985) 162–163. https://doi.org/10.1038/318162a0.
[2] Y. Zhang, Y.W. Tan, H.L. Stormer, P. Kim, Experimental observation of the quantum Hall effect and Berry's phase in graphene, Nature 2005 438:7065 438 (2005) 201–204. https://doi.org/10.1038/nature04235.
[3] J.W.F. To, Z. Chen, H. Yao, J. He, K. Kim, H.H. Chou, L. Pan, J. Wilcox, Y. Cui, Z. Bao, Ultrahigh Surface Area Three-Dimensional Porous Graphitic Carbon from Conjugated Polymeric Molecular Framework, ACS Cent Sci 1 (2015) 68–76. https://doi.org/10.1021/ACSCENTSCI.5B00149.
[4] V. Georgakilas, J.A. Perman, J. Tucek, R. Zboril, Broad Family of Carbon Nanoallotropes: Classification, Chemistry, and Applications of Fullerenes, Carbon Dots, Nanotubes, Graphene, Nanodiamonds, and Combined Superstructures, Chem Rev 115 (2015) 4744–4822. https://doi.org/10.1021/CR500304F.

[5] G. Lalwani, A. Gopalan, M. D'Agati, J. Srinivas Sankaran, S. Judex, Y.X. Qin, B. Sitharaman, Porous Three-Dimensional Carbon Nanotube Scaffolds for Tissue Engineering, J Biomed Mater Res A 103 (2015) 3212. https://doi.org/10.1002/JBM.A.35449.

[6] H. Singh, A.B. Santos, D. Das, R.S. Ambekar, P. Saxena, C.F. Woellner, N.K. Katiyar, C.S. Tiwary, Stress concentration targeted reinforcement using multi-material based 3D printing, Appl Mater Today 36 (2024) 102010. https://doi.org/10.1016/J.APMT.2023.102010.

[7] S.M. Sajadi, P.S. Owuor, S. Schara, C.F. Woellner, V. Rodrigues, R. Vajtai, J. Lou, D.S. Galvão, C.S. Tiwary, P.M. Ajayan, Multiscale Geometric Design Principles Applied to 3D Printed Schwarzites, Advanced Materials 30 (2018) 1704820. https://doi.org/10.1002/ADMA.201704820;REQUESTEDJOURNAL:JOURNAL:15214095;WGROUP:STRING:PUBLICATION.

[8] L.C. Felix, R.M. Tromer, C.F. Woellner, C.S. Tiwary, D.S. Galvao, Mechanical response of pentadiamond: A DFT and molecular dynamics study, Physica B Condens Matter 629 (2022) 413576. https://doi.org/10.1016/J.PHYSB.2021.413576.

[9] B. Mortazavi, F. Shojaei, X. Zhuang, L.F.C. Pereira, First-principles investigation of electronic, optical, mechanical and heat transport properties of pentadiamond: A comparison with diamond, Carbon Trends 3 (2021) 100036. https://doi.org/10.1016/J.CARTRE.2021.100036.

[10] K.I. Tserpes, P. Papanikos, Continuum modeling of carbon nanotube-based super-structures, Compos Struct 91 (2009) 131–137. https://doi.org/10.1016/J.COMPSTRUCT.2009.04.039.

[11] L.J. Gibson, M.F. Ashby, Cellular solids: Structure and properties, second edition, Cellular Solids: Structure and Properties, Second Edition (2014) 1–510. https://doi.org/10.1017/CBO9781139878326.

[12] J. Bauer, L.R. Meza, T.A. Schaedler, R. Schwaiger, X. Zheng, L. Valdevit, Nanolattices: An Emerging Class of Mechanical Metamaterials, Advanced Materials 29 (2017) 1701850. https://doi.org/10.1002/ADMA.201701850;REQUESTEDJOURNAL:JOURNAL:15214095;WGROUP:STRING:PUBLICATION.

[13] Y. Yang, J. Xu, Y. Chen, Y. Xia, C. Schäfer, M. Ratsch, M. Rahm, T. Willhammar, K. Börjesson, Diamondiyne: A 3D carbon allotrope with mixed valence hybridization, (2025). https://doi.org/10.26434/CHEMRXIV-2025-1B0MV.

[14] C.M. de O. Bastos, E.J.A. dos Santos, R.A.F. Alves, A. Cavalheiro Dias, L.A. Ribeiro Junior, D.S. Galvão, Entangled Interlocked Diamond-like (Diamondiynes) Lattices, ACS Omega 10 (2025) 46065–46070. https://doi.org/10.1021/ACSOMEGA.5C07159/SUPPL_FILE/AO5C07159_SI_011.MP4.

[15] R.H. Baughman, D.S. Galvão, Tubulanes: carbon phases based on cross-linked fullerene tubules, Chem Phys Lett 211 (1993) 110–118. https://doi.org/10.1016/0009-2614(93)80059-X.

[16] A.K. Singh, A.K. Nath, D.K. Pratihar, A. Roy Choudhury, Development and performance of additively manufactured Tungsten carbide tool insert: An application of additive manufacturing towards flexible and customized machining, Tribol Int 197 (2024) 109780. https://doi.org/10.1016/J.TRIBOINT.2024.109780.

[17] D. Zhao, Z. Zhou, K. Ruan, X. Xu, G. Zhang, Z. Chen, K. Wang, Y. Xiong, In-process density measurement for model-based process optimization of functionally graded foam microcellular structures in material extrusion additive manufacturing, Addit Manuf 106 (2025) 104817. https://doi.org/10.1016/J.ADDMA.2025.104817.

[18] A.K. Singh, A. Mahata, A.K. Nath, D.K. Pratihar, A. Roy Choudhury, Thermal stress and crack mitigation in additive manufacturing of tungsten carbide with preheating: An application of analytical modeling with real-time thermal signal feedback, J Manuf Process 148 (2025) 330–344. https://doi.org/10.1016/J.JMAPRO.2025.05.050.

[19] A. Cano-Vicent, M.M. Tambuwala, S.S. Hassan, D. Barh, A.A.A. Aljabali, M. Birkett, A. Arjunan, Á. Serrano-Aroca, Fused deposition modelling: Current status, methodology, applications and future prospects, Addit Manuf 47 (2021) 102378. https://doi.org/10.1016/J.ADDMA.2021.102378.

[20] H. Singh, L. V. Bastos, D. Das, R.S. Ambekar, C. Woellner, N.M. Pugno, C.S. Tiwary, Composite strengthening via stress-concentration regions softening: The proof of concept with

Schwarzites and Schwarzynes inspired multi-material additive manufacturing, Addit Manuf 90 (2024) 104336. https://doi.org/10.1016/J.ADDMA.2024.104336.

[21] R.S. Ambekar, E.F. Oliveira, P. Pugazhenthi, S. Singh, D. Roy Mahapatra, D.S. Galvao, C.S. Tiwary, Impact Resistance of Complex 3D-Printed Schwarzites Structures, Adv Eng Mater (2025) e202502191. https://doi.org/10.1002/ADEM.202502191;CSUBTYPE:STRING:AHEAD.

[22] H. Singh, A.B. Santos, D. Das, R.S. Ambekar, C.F. Woellner, S. Nagarajaiah, C.S. Tiwary, Zeolites topology inspired multi-material-based 3D printing of porous composite structures with high resilience, Progress in Additive Manufacturing 2025 10:10 10 (2025) 8157–8169. https://doi.org/10.1007/S40964-025-01103-7.

[23] L.C. Felix, V. Gaál, C.F. Woellner, V. Rodrigues, D.S. Galvao, Mechanical Properties of Diamond Schwarzites: From Molecular Dynamics Simulations to 3D Printing, (2020). https://arxiv.org/pdf/2006.02848 (accessed December 21, 2025).

[24] A.C.T. Van Duin, S. Dasgupta, F. Lorant, W.A. Goddard, ReaxFF: A Reactive Force Field for Hydrocarbons, Journal of Physical Chemistry A 105 (2001) 9396–9409. https://doi.org/10.1021/JP004368U.

[25] R.S. Ambekar, E.F. Oliveira, B. Kushwaha, V. Pal, L.D. Machado, S.M. Sajadi, R.H. Baughman, P.M. Ajayan, A.K. Roy, D.S. Galvao, C.S. Tiwary, On the mechanical properties of atomic and 3D printed zeolite-templated carbon nanotube networks, Addit Manuf 37 (2021) 101628. https://doi.org/10.1016/J.ADDMA.2020.101628.

[26] S. Nosé, A unified formulation of the constant temperature molecular dynamics methods, J Chem Phys 81 (1984) 511–519. https://doi.org/10.1063/1.447334.

[27] W. Humphrey, A. Dalke, K. Schulten, VMD: Visual molecular dynamics, J Mol Graph 14 (1996) 33–38. https://doi.org/10.1016/0263-7855(96)00018-5.

[28] H. Singh, R.S. Ambekar, D. Das, V.A. Danam, N.K. Katiyar, B. Kanti Das, C.S. Tiwary, J. Bhattacharya, Enhancing structural resilience by using 3D printed complex polymer reinforcement for high damage tolerant structures, Constr Build Mater 425 (2024) 136085. https://doi.org/10.1016/J.CONBUILDMAT.2024.136085.